\documentclass[12pt,preprint]{aastex}
\usepackage{graphics}
\input epsf
\begin{document}
\title{A Fractal Origin for the Mass Spectrum of Interstellar Clouds:
II. Cloud Models and Power Law Slopes}
\author{Bruce G. Elmegreen
  \affil{IBM Research Division, T.J. Watson Research Center,
    P.O. Box 218, Yorktown Heights, NY 10598, USA}
}

\begin{abstract} Three-dimensional fractal models on grids of $\sim200^3$
pixels are generated from the inverse Fourier transform of noise with
a power law cutoff, exponentiated to give a log normal distribution
of density.  The fractals are clipped at various intensity levels and
the mass and size distribution functions of the clipped peaks and their
subpeaks are determined. These distribution functions are analogous to the
cloud mass functions determined from maps of the fractal interstellar
medium using various thresholds for the definition of a cloud. The
model mass functions are found to be power laws with powers ranging
from $-1.6$ to $-2.4$ in linear mass intervals as the clipping level
increases from $\sim0.03$ to $\sim0.3$ of the peak intensity.  The low
clipping value gives a cloud filling factor of $\sim10$\% and should
be a good model for molecular cloud surveys.  The agreement between
the mass spectrum of this model and the observed cloud and clump mass
spectra suggests that a pervasively fractal interstellar medium can be
interpreted as a cloud/intercloud medium if the peaks of the fractal
intensity distribution are taken to be clouds. Their mass function is a
power law even though the density distribution function in the gas is a
log-normal. This is because the size distribution function of the clipped
clouds is a power law, and with clipping, each cloud has about the same
average density.  A similar result would apply to projected clouds that
are clipped fractals, giving nearly constant column densities for power
law mass functions.  The steepening of the mass function for higher clip
values suggests a partial explanation for the steeper slope of the mass
functions for star clusters and OB associations, which sample denser
regions of interstellar gas.  The mass function of the highest peaks is
similar to the Salpeter IMF, suggesting again that stellar masses may
be determined in part by the geometry of turbulent gas.  \end{abstract}

\keywords{turbulence --- ISM: clouds --- 
ISM: structure --- open clusters and associations: general }

\section{Introduction} 

Interstellar gas appears scale-free when viewed with Fourier transform
power spectra (Crovisier \& Dickey 1983; Green 1993; Lazarian \& Pogosyan
2000; St\"utzki et al. 1998; Stanimirovic et al. 1999; Elmegreen, Kim, \&
Staveley-Smith 2001), delta variance techniques (St\"utzki et al. 1998;
Zielinsky \& St\"utzki 1999), spectral correlation functions (Rosolowsky
et al.  1999), principal component analysis (Heyer \& Schloerb 1997),
perimeter-area measures (Dickman, Horvath, \& Margulis 1990; Falgarone,
Phillips, \& Walker 1991), box-counting techniques (Westpfahl et al.
1999), and multifractal analysis (Chappell \& Scalo 2001).

The interstellar medium (ISM) looks like a collection of discrete clouds,
however, when intensity contours are drawn (Solomon et al. 1987; Loren
1989; St\"utzki \& G\"usten 1990; Williams, de Geus, \& Blitz 1994;
Lemme et al. 1995; Kramer et al. 1998; Heyer, Carpenter, \& Snell 2001)
or when spectral absorption lines are fit to Gaussians (Adams 1949; Hobbs
1978; Clark 1965; Radhakrishnan \& Goss 1972; Spitzer \& Jenkins 1975).

These two interpretations have led to distinct models for the origin of
gas structure and star formation. Scale-free models typically involve
turbulence and self-gravity (Falgarone \& Phillips 1990; Scalo 1990;
Pfenniger \& Combes 1994; Lazarian 1995; Elmegreen 1999b; Rosolowsky
et al. 1999; MacLow \& Ossenkopf 2000; Pichardo et al. 2000; Klessen,
Heitsch, \& MacLow 2000; V\'azquez-Semadeni, Gazol, \& Scalo 2000; Semelin
\& Combes 2000; Ostriker, Stone, \& Gammie 2001; Heitsch, Mac Low, \&
Klessen 2001; Toomre \& Kalnajs 1991; Wada \& Norman 1999, 2001).

Cloudy models involve sticky collisions and star formation triggered by
colliding and compressed clouds (Kwan 1979; Hunter et al. 1986; Tan 2000;
Scoville, Sanders, \& Clemens 1986).

For general interstellar gas dynamics, the turbulent model (von Weizsacker
1951; Sasao 1973) may be more realistic than the cloudy model for the
origin of structure (LaRosa, Shore \& Magnani 1999; Ballesteros-Paredes,
Hartmann, \& V\'azquez-Semadeni 1999). Turbulence also gives cloud-like
spectral lines through both density structure and velocity crowding
(Ballesteros-Paredes, V\'azquez-Semadeni, \& Scalo 1999; Lazarian \&
Pogosyan 2000; Pichardo, et al. 2000).

For sudden transition fronts like expanding shells, spiral arms, and dust
lanes, cloud collisions may be an appropriate way to model the dynamics
(e.g., Kenney \& Lord 1991; Elmegreen 1988). This is because the pre-front
clumps made by turbulence at slow ambient speeds are forced to collide
together and interact at much faster speeds inside the front. Strongly
self-gravitating clouds that are made by turbulence at moderate speeds
but then given a chance to cool and settle to high densities should also
interact as in the cloudy models. Because of their high densities, these
clouds or clumps should move somewhat independently of the surrounding
turbulent gas, perhaps lagging behind the expanding flows to make bright
rims, or punching through spiral dustlanes to make feathery structures
(Seth 2000).

The most obvious point of contact between these two views of gas structure
is the mass spectrum of the regions that are isolated enough to be defined
as clouds. The scale-free nature of the gas shows up as a scale-free
mass spectrum for the clouds (Elmegreen \& Falgarone 1996; St\"utzki et
al. 1997). This mass spectrum is not useful in the turbulent model because
it does not reflect the continuous distribution of matter that is really
present. The spectrum is more important for bound star clusters, which
are better defined (Elmegreen et al. 2000), and for individual stars,
whose masses are probably proportional to the primordial clump masses
(Motte, Andr\'e, \& Neri 1998; Testi \& Sargent 1998; Bacmann et al.
2000; Tachihara et al. 2000).

Here we model the cloud mass and size spectra using the intensity peaks in
a simulated fractal to represent clouds. The results show the anticipated
$\sim M^{-2}dM$ spectrum that comes from simple hierarchical models
(Fleck 1996), but for a different reason than what is usually given. In
the usual interpretation, the $M^{-2}dM$ spectrum comes from the fact
that there is a constant total mass in each logarithmic interval of mass;
i.e., each small clump at the bottom is contained inside each large clump
at the top. Then $M\xi(\log M)d\log M=$ constant$\times d\log M$ becomes
$n(M)dM\equiv\xi(\log M)d\log M= \xi(\log M)dM/M\propto M^{-2}dM$. We
show here that the bottom of the hierarchy, where the gas density has
local isolated peaks, has about the same spectrum. This is true even
though the probability distribution function of density alone is not a
power law, but log-normal.

We also find that the spectrum flattens to $M^{-1.6}dM$ as the lower
limit to the cloud intensity is decreased. This flattening may explain
the difference between the mass spectrum of star clusters, which is
close to $M^{-2}dM$ (Battinelli et al. 1994; Elmegreen \& Efremov 1997;
Whitmore \& Schweizer 1995; Zhang \& Fall 1999) and the mass spectrum
for clouds, which typically ranges between $M^{-1.5}$ and $M^{-1.8}$
(e.g., Blitz 1993; Kramer et al. 1998; Heyer, Carpenter, \& Snell 2001).
An even steeper spectrum results from the highest clipping levels modeled
here and is close to the Salpeter IMF for stars.  Thus the increase
in slope of the mass functions from clouds to clusters to stars may be
partly the result of different density thresholds in the same fractal gas.

\section{Models}

Fractal Brownian motion clouds (Stutzki et al. 1998) are
generated by first filling a 3D lattice in wavenumber space,
$(k_x,k_y,k_z)$, with noise distributed as a Gaussian with a
dispersion of unity. The noise cube is multiplied by $k^{-5/3}$ for
$k=\left(k_x^2+k_y^2+k_z^2\right)^{1/2}$. The inverse three-dimensional
FFT of the resulting truncated noise cube gives a fractal (Voss
1988) with a Gaussian distribution of intensity, $I_0(x,y,z)$, as
shown on the bottom of Figure 1. To simulate a turbulent fractal, we
exponentiate this intensity distribution, $I(x,y,z)=\exp\left(\alpha
I_0(x,y,z)/I_{0,max}\right)$ for maximum original intensity $I_{0,max}$
and contrast factor $\alpha=4$. This gives another fractal, now with a
log-normal density distribution (Fig. 1, top). The motivation for the
log-normal comes from studies by V\'azquez-Semadeni (1994), Nordlund \&
Padoan (1999), Klessen 2000, and Wada \& Norman (2001).

A fractal obtained in this way is a continuous distribution of density and
more properly called a multifractal because the local fractal dimension of
the density structure varies from the peaks to the valleys (e.g., Chappell
\& Scalo 2001; Vavrek 2001). We refer to it here only as a fractal.

For the cloud mass and size spectra determined here, we consider two
cloud models. In the first (Sect. \ref{sect:model1}), a cloud is defined
as all of the emission above a fixed cutoff in density. In the second
(Sect. \ref{sect:model2}), clouds that are resolved as separate peaks
are each counted, whether or not they occur inside a broader emission
region above the cutoff.  The broader emission in this second case
is divided into subparts to go with each peak. The subpart masses are
taken proportional to the peak masses that were previously determined
from a higher cutoff level.  The algorithm will be described in more
detail later.

For the first set of models, the fractal density distribution in which the
clouds were counted was taken to be the inner $180^3$ pixels of a $210^3$
grid, to avoid edge effects.  In the second case, the inner $160^3$
px inside a $192^3$ px grid was used.  Models with many different grid
sizes were made to confirm that the basic results are independent of
this. Larger grids have more clouds; the first model had 73,000 separate
clouds above 0.03 times the peak density, and the second model had
40,000 clouds and subclouds above $e^{-3}=0.05$ times the peak density.
The ratio of peak to minimum value in the density distribution was 2400
in the first set of models, and 3800 in the second set; this ratio is
determined by the $\alpha$ value.

\subsection{Model 1: 
Clouds as Emission Regions above a Fixed Density Cutoff}
\label{sect:model1}

The central $100^3$ pixels of the first model is shown in Figure 2,
clipped to display everything brighter than 0.03 of the peak.  This
clipping level gives the most reasonable agreement with interstellar
cloud surveys, as discussed below.

To find clouds in the $180^3$ px$^3$ fractal, we begin by initiating
another $180^3$ cube, $J(x,y,z)$, with values all equal to 0.  This cube
keeps track of which pixels in the fractal cube have been counted in the
mass sum. Then, to find a cloud, we begin $10$ pixels inside each edge
in one corner of the fractal cube and step along in one direction in a
regular fashion until a pixel value is reached that exceeds the clipping
threshold. We determine whether this pixel has been counted before by
viewing the value of the corresponding pixel in $J(x,y,z)$. If $J=0$,
then it has not been counted before and we proceed to count the cloud
mass. If $J=1$, we continue with our regular search for another cloud. In
the case where $J=0$ for this pixel, we add the density of the pixel
to the total mass of this cloud, which begins at 0, and we change the
corresponding pixel in the $J$ cube from 0 to 1. We then begin to map
around inside this cloud by stepping one pixel away from the current
pixel in a random direction. If the new pixel has a value lower than
the threshold, we have just crossed the cloud boundary so we go back. If
the new pixel value is greater than the threshold so we are still inside
the cloud, then we determine if this position has already been added to
the cloud mass sum by reading the value of $J$ at this position. If the
new pixel has not been counted in the current cloud (i.e., if the $J$
value is still 0), then we add its density to the running sum for that
cloud and set the corresponding value in $J$ to 1.

This procedure continues until further random searching returns no
new pixels in the same cloud after a total number of additional tries
equal to $4N^2$, where $N$ is the number of pixels already counted
for that cloud.  If $N$ is small or large, then the minimum or maximum
trial counts were taken equal to 1000 and $10^7$.  This step number
requirement gives the random walk among pixels a reasonable chance to
visit every pixel inside the cloud, even if the boundary is very ragged
and there are cul de sacs and spikes which would nearly trap a shorter
random search. When the cloud mass is summed, we add the result to a
histogram of cloud masses, which is the mass spectrum, using a linear
interval of the mass for counting.  The cloud search then continues
along the regular search path where it was before the most recent cloud
was mapped.  Cloud searches are much faster for higher intensity cut-off
levels because the clouds are smaller and the search requirement of 
$4N^2$ steps without a new cloud pixel is more easily reached.

The clouds modeled here are solid objects, not projected objects.
We consider cloud structure in the full three-dimensional grid because
spectral line mapping in real data can usually distinguish between
different three-dimensional objects at different velocities on the same
line of sight.  Thus the fractal dimension of observed interstellar CO
clouds and clumps, obtained from the size distribution of contour-mapped
objects, was found to be around 2.3 -- one more than the projected fractal
dimension of 1.3 obtained from the boundary structure (Pfenniger \&
Combes 1994; Elmegreen \& Falgarone 1997).  This approximation does not
account for artificial clouds that may appear in real surveys because
of velocity crowding in turbulent gas (Lazarian \& Pogosyan 2000).

Figure 3 shows mass spectra for intensity clipping values equal to
0.03, 0.1, and 0.3 times the peak intensity.  The spectrum for the 0.3
clipping level is an average of the spectra for 10 different random
fractal models; this averaging was done in order to reduce the noise
in the cloud counting, considering the small number of bright peaks in
each model.  The spectrum for the 0.1 clipping level is an average over
4 random fractal models, and the spectrum for 0.03 is an average over 2
fractal models.  The spectral slopes are approximately $\alpha=-1.7$,
$-1.9$ and $-2.3$, respectively. These correspond to power law mass
functions $n(M)dM\propto M^{\alpha}dM$ in linear intervals of mass, or
mass functions $\propto M^{1+\alpha}d\log M$ in logarithmic intervals
of mass.

Figure 4 shows the running average slopes of the mass functions versus the
cloud mass for different clipping levels. These average slopes start at
the lowest cloud mass for that clipping level and extend up to the mass
$M$ plotted on the abscissa. The slopes of the power law mass spectra
become inaccurate at high mass because the number of clouds goes to zero
and the individual spectra level off. At low to intermediate mass, the
slopes are about constant for a range in cloud mass that spans a factor
of $\sim100$ for low clipping levels. This is the mass range where the
spectrum is a power law. The range is smaller for higher clipping levels
because the cloud masses are smaller.

The slope of the power law mass spectrum gets steeper with higher
clipping levels.  To understand the origin of this increase, we ran
other models with smaller peak-to-valley density contrasts  (e.g., using
$\alpha=1.4$). The same range in clump spectral slopes resulted for the
high and low clipping levels, but now the range in clipping levels was
only a factor of 5 because of the smaller density contrast for the low
$\alpha$ case, rather than 30 as in the Figures.  This change suggests
that the mass spectrum slope decreases for low clipping values in both
cases because the low density limit of the fractal is approached. Near
the low density limit, each cloud has a sprawling boundary that spans
nearly the distance to the next cloud, and there is a higher proportion
of high mass clouds. There is also very little intercloud medium at low
clipping levels. For the lowest clipping level, 0.03 times the peak, the
threshold cloud density is $\sim70$ times the minimum fractal density,
which is $\sim1/2400$ of the peak.  This is still far from the lowest
density level, but apparently close enough to change the geometric
characteristics.

Figure 5 shows the size distribution function for the model clipped at
0.03 times the peak value.  The cloud size is taken to be the cube-root of
the number of pixels.  The size distribution function is approximately a
power law with a slope of $\sim-3.75$.  This plot is again a histogram,
determined for linear intervals of size. Thus the average fractal
dimension is one more than this slope, or $D\sim-2.75$ (Mandelbrot 1982).
This is not a very meaningful concept for a multifractal with a smooth
distribution of density because there is a range of fractal dimensions
depending on location, but the result is analogous to the dimension
determined from the size distribution of interstellar clouds, which gives
about $D\sim-2.3$ (Elmegreen \& Falgarone 1996).  The size spectra for
the models with other clipping factors are not shown because the size
ranges are too small to be interesting; they tend to be steeper for
higher clipping levels like the mass spectra.

Figure 6 shows the mass versus size for all clouds.  The regressions
with slopes of $\sim3.3$ indicate that all clouds found with a particular
clipping level have about the same average density and that the density
scales in proportion to the clipping level.  This is to be expected
because most of the cloud mass is on the periphery of the cloud, where the
volume is greatest, and the density there is about the clipping value.
This cubic dependence is similar to that found for clump studies inside
molecular clouds (Loren 1989; St\"utzki \& G\"usten 1990; Williams, de
Geus, \& Blitz 1994; Williams, Blitz, \& Stark 1995; see the summary of
observations in Elmegreen \& Falgarone 1996) and differs significantly
from the quadratic dependence found by Larson (1981).

The difference in the mass-radius relations for clump surveys and whole
GMC surveys seems to be related to the threshold used for the observation.
In the clump case, the threshold is probably the critical density for
excitation of the molecule (as recognized, for example, by Williams,
Blitz, \& Stark 1995). In Larson's correlation, which was based on large
scale CO surveys and later confirmed by other whole-cloud surveys (e.g.,
Solomon et al.  1987), the threshold is CO column density, probably tied
to the telescope sensitivity.  For example, the Solomon et al. survey
used a brightness temperature limit to define CO clouds.  Thus the
fundamental $M\propto R^2$ result of Larson (1981), which, when combined
with the virial theorem and Kolmogorov law, is the basis for most of the
correlations that have been found for molecular clouds, seems to be an
artifact of sampling near the telescope limit of detection, which is a
column density limit for a fractal cloud (see also V\'azquez-Semadeni,
Ballesteros-Paredes, \& Rodriguez, 1997).

\subsection{Model 2: Clouds and Subclouds Defined by Each Peak}
\label{sect:model2}

In the second model, clouds were defined in a $160^3$ px$^3$ grid
by local peaks in the density, whether or not they were separated by
a region below the current density cutoff.  This is analogous to the
definition of a cloud or clump in the surveys by Williams et al. (1994,
1995) and St\"utzki, \& G\"usten (1990), who fit the peaks and their
surrounding emission with various strategies.  This model differs from
the first model discussed above because a single cloud in the first case
could be divided into several sub-clouds in the second case if there are
separate peaks.  To keep track of where the peaks are, we used a third
cube, $P(x,y,z)$, of size $160^3$ with entries equal to the peak number.
A second cube, $J(x,y,z)$ as in model 1, kept track of which pixels had
been searched already.

The peaks in the density are found by successively lowering the threshold,
which begins at $\exp(-1)$ times the overall peak density in the fractal
cube, and then goes to $\exp(-2)$ and $\exp(-3)$ times this peak density.
The peaks found with the $\exp(-1)$ threshold are located in the
same way as for model 1, with no separate subpeaks. Presumably they
would be unresolved also in a real cloud if the peak intensity stands
above the detection threshold by a factor less than $e^1=2.7$.
These first-threshold clouds are then used by the program again, but now
with the threshold of $\exp(-2)$.  When a new cloud was found standing
above this $\exp(-2)$ threshold, the peak-identification cube, $P$,
was viewed at all of the positions belonging to this new cloud to see
if there were already any peaks in it. If there were no peaks, then the
current cloud was defined to be a new peak and its full mass was added
to the list of peaks. If there was one previous peak, then the mass from
this level was given to the previous peak.  If there was more than one
previous peak from the next higher cutoff level, then the mass of the
current cloud was divided up among the higher peaks.  With this strategy,
a distinct peak is a region separated from another peak by a level of
emission that dips below $\sim\exp\left(-1\right)$ of the peak value.

This partitioning of mass among the various peaks in a threshold-defined
cloud was done in several different ways.  In one method, the peaks
were fit to Gaussians when they were first found, and the extrapolated
masses from these Gaussians were determined at the next lower level
when they appeared inside a cloud there.  This mass at the next lower
level was then divided up among the former peaks in proportion to their
extrapolated Gaussian masses.  In another method, the current cloud was
divided into pieces proportional to the former peak masses themselves,
without any Gaussian extrapolations.  The difference between these two
partitionings was noticeable but not large.  Because the peak masses
from the levels where they are first discovered are distributed as a
power law, as shown in the previous subsection, any partitioning of
the current cloud in proportion to these previous peaks maintains that
same power law and extends it to larger masses.  Thus the combined mass
distribution of isolated peaks and subdivided clouds is similar to that
of the peaks alone.  In what follows, we discuss this case, i.e., where
the partitioning is proportional to the previous peak masses.  In the
Gaussian-fitting case, the extrapolated Gaussian masses could be much
larger than the pure peak masses if the Gaussian dispersions were large,
and this led to a wild variation in how the clouds in the current level
were divided among peaks from the higher levels.  It gave results that
were similar to a third case where the current cloud was equally divided
among the peaks from the higher level.

In the final spectrum down to the third level ($e^{-3}$), the Gaussian
and equal-division methods did not produce a continuous power law, but
had a power law at low mass, where the peaks were single, and a broad
concentration of masses at intermediate values, where the subdivisions
gave a somewhat random mixture of masses.  The preferred case, with the
subdivisions directly proportional to the former peak masses, gave an
approximately continuous power law.

Figure 7 shows the mass distribution functions for clouds and sub-clouds
defined in this way using density cutoffs of $\exp(-1)$ (on the left),
$\exp(-2)$, and $\exp(-3)$ (on the right).  The top three panels have
no added noise, the bottom three have noise at the level of $0.5e^{-3}$
times the peak density.  This noise affects only the mass spectrum made
from the lowest contour (plotted on the right), and it tends to add
low mass clouds, which are mostly noise.  The average slope of the mass
spectrum does not change much with added noise.

The dashed lines that lie nearly parallel to the distribution functions
in Figure 7 have slopes of $-2.4$, $-1.8$, and $-1.6$ for these three
cutoffs, respectively.  These slopes are similar to the mass spectrum
slopes found in the previous section for the cases with similar density
cutoffs.  Thus the division of big clouds into subclouds does not affect
the spectrum much. There is a tendency for the mass spectra to become
slightly steeper at low mass, which is the result of the steepening
effect found previously for small clouds at high cutoff levels: most
of these low mass clouds are single peaks near the threshold.  Figure 8
shows the running average slopes for the noise-free cloud/subcloud case,
as in Figure 4.  The other figures shown in section \ref{sect:model1},
namely, the mass-size relation and the size spectrum, are essentially
the same for the present case.

\section{Discussion}

The connection between the cloud model of interstellar structure and
the multifractal model is readily understood when clouds are viewed as
isolated peaks in the fractal. Here we have shown that the mass spectrum
for such fractal clouds is similar to the observed cloud spectrum when
clouds are defined or selected to be those regions where the local density
exceeds several percent of the peak (resolved) density, or when clouds
are defined to be the resolvable peaks plus a proportional amount of gas
in the underlying plateaus.  Unresolved peaks can have higher densities,
as can a fractal model with more cells, but this will not affect the
mass spectrum of the resolved objects.

Observational cloud surveys tend to span only a factor of 5 to 10 in
physical scale (see references for cloud and clump surveys given above).
This limited range is entirely a selection effect because unbiased
surveys analyzed with power spectra or similar techniques find a much
wider range of scales in the interstellar medium, up to a factor greater
than $\sim100$ in the case of whole galaxies. In the cloud surveys, the
factor of $\sim10$ limit arises because smaller regions are unresolved and
larger regions are subdivided into separate clouds, the whole structure
being hierarchical (Scalo 1985).

For many CO surveys, a factor of 10 in scale corresponds to a factor
of 10 in average density (Larson 1981), and a factor of 10 in density
corresponds to a volume filling factor of 10\%. This latter result is
because for an infinitely self-similar fractal, the filling factor scales
inversely with the density (Elmegreen 1999c). In fact, most cloud surveys
report a filling factor of about $\sim10$\% (e.g., see review in Blitz
1993); this is only an artifact of sampling in the fractal point of view.
In the current models, the average filling factors for the clouds in
Section \ref{sect:model1} are $1.2\times10^{-4}$, $7.0\times10^{-3}$,
and 0.13 for clipping densities of 0.3, 0.1, and 0.03 times the peak.
For the clouds in section \ref{sect:model2}, they are $1.3\times10^{-4}$,
$5.7\times10^{-3}$, and 0.065.

These filling factors vary approximately as the inverse cube or
inverse square of the density threshold because the low clipping levels
approach the minimum density, and the sprawling cloud boundaries take
up excess volume.  The models with the lowest clipping levels, 0.03 and
$e^{-3}=0.05$, have the about same filling factors as the observed clouds
and clumps in surveys.

The point here is that the density range that is inadvertently selected
for real cloud surveys is also what the fractal models need to give
the observed filling factor and mass function. If different observing
techniques are employed, giving a wider range of cloud sizes or average
densities, for example, thereby increasing the clump filling factor, then
we predict that the slope of the resulting mass function will decrease.
Similarly, surveys with molecular tracers sensitive to very high densities
should be selecting regions closer to the peaks of the turbulent fractal,
with lower filling factors, and these surveys should get slightly steeper
slopes for the clump mass spectrum.

Note that the absolute density of an observed region is not important for
the slope of the mass spectrum, only the relative density compared to the
peaks and valleys in that region. As the threshold density approaches
the minimum density, the mass spectrum flattens because the cloud
boundaries spread out and the low density "intercloud" medium, which
is just the sub-threshold gas in the same fractal, gets included more
and more with each cloud. This interpretation is valid for molecular
clouds and other gases with a single phase of thermal temperature.
Of course, the intercloud means something different when there is a
high temperature phase.  For example, fractal models of the neutral
hydrogen in the Large Magellanic Cloud require a multi-phase structure
because of the high density contrasts that are present (Elmegreen, Kim,
\& Staveley-Smith 2001).

The size distribution function for our model clouds is also similar
to that for interstellar clouds in the best-fit case.  The power law
nature of the size spectrum, combined with the cutoff in density in the
definition of a cloud, ensures that the mass spectrum is also a power law.
This is because the average density is always some factor of order unity
times the cutoff. Thus the mass spectrum is a power law even though the
density distribution function is a log-normal.

The mass spectrum for star clusters is steeper than that for clumps,
by several tenths (cf. Sect. 1). The fractal model suggests that this
steepness results from the denser cloud regions that are sampled by
clusters. The typical density of an embedded star cluster corresponds
to an H$_2$ density of around $10^5$ cm$^{-3}$ (Lada, Evans \& Falgarone
1997). This is much closer to the peak density of the turbulent fractal
than the density threshold for the cloud itself, so the mass spectrum
should be steeper in this model. However, gravity modifies the gas density
in the region of a cluster, so the present results are only suggestive.

The densest peaks sampled here, with $>30$\% of the peak density in
the cloud, have mass spectra with slopes of $\sim-2.3$ and $\sim-2.4$
for models 1 and 2, respectively. These are similar to the slope of
the Salpeter stellar IMF.  Indeed, mm-continuum sources and stars
do sample denser regions than clusters, so the increase in slope is
sensible. However, other factors enter into the stellar IMF, such as
the relative rate of collapse in regions with different masses and the
competition for mass. Thus the importance of the purely fractal result
for the IMF cannot be assessed without further modeling (see Elmegreen
1997, 1999a).

\section{Summary}

Clouds that are defined by the peaks of a smooth fractal distribution,
whether deliberately or inadvertently because of the observing procedure,
have power law mass distributions even if the gas density has a
log-normal distribution. The power of the mass distribution ranges
from $-1.6$ to $-2.4$ as the clipping level varies from 0.03 to 0.3
times the peak. The shallower power is similar to what is observed for
interstellar clouds, and the cloud filling factor in the model is similar
to the observations too.  The steeper power laws are similar to what
are observed for star clusters and mm-continuum sources, respectively,
perhaps because clusters and continuum sources sample greater densities
than the outer cloud boundary.

The origin of the power law for the cloud mass spectrum is the power law
for the cloud size spectrum at a fixed threshold in density or column
density, depending on the nature of the survey.  Many of the correlations
observed for molecular clouds and clumps now appear to be artifacts of
these density and column density thresholds, considering the fractal
nature of the interstellar medium and the internal correlations from
turbulent motions.

Acknowledgements:
This work was supported by NSF Grant AST-9870112 to B.G.E.  Helpful comments
by the referee led to the second clump-finding algorithm discussed here.

\clearpage

\begin{figure}
\plotone{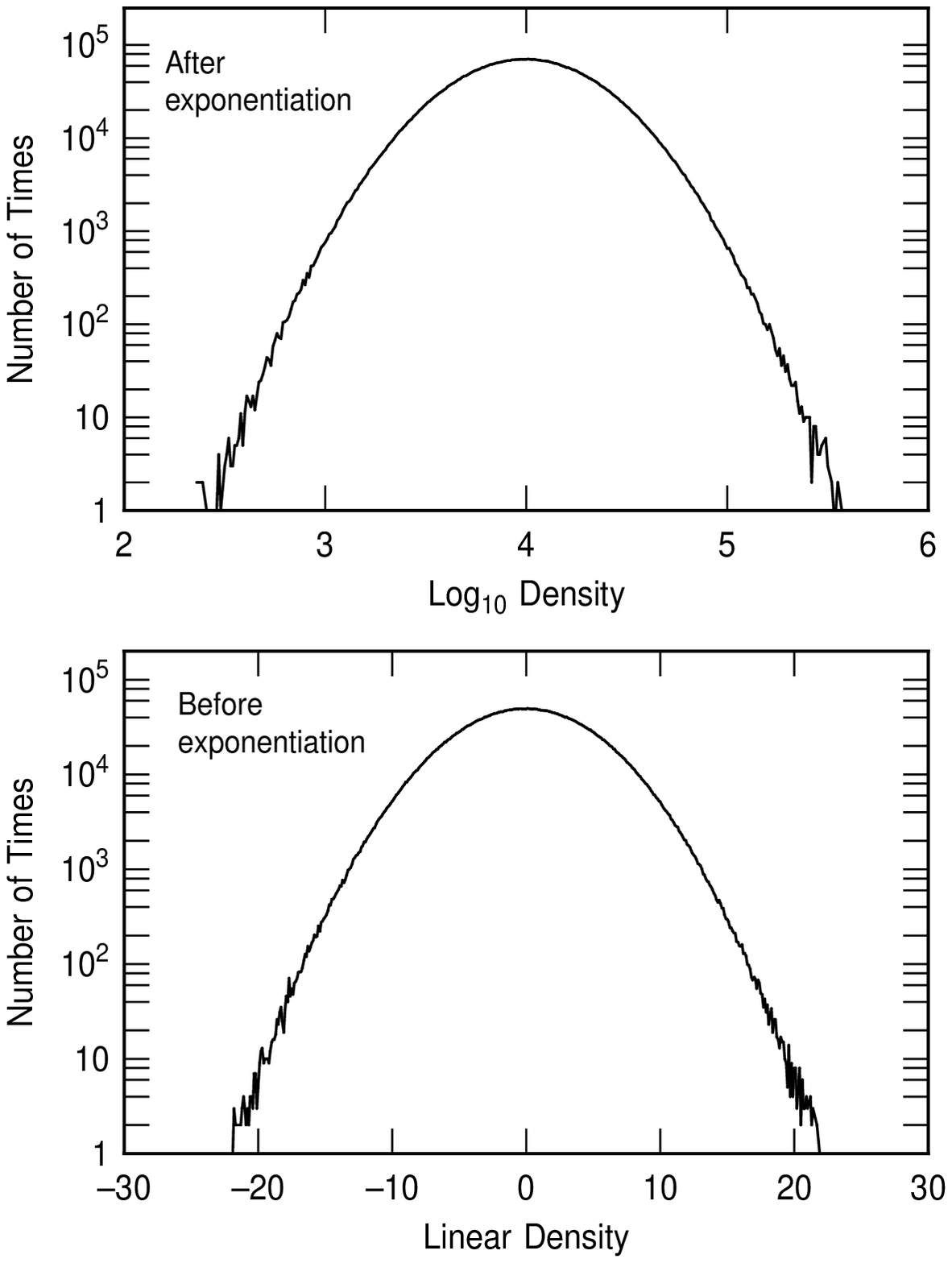}
\caption{(bottom) Histogram of density in the fractal model that 
was made
from the inverse Fourier transform of noise with a power
law cut off. (top) Histogram of the fractal model used for the
cloud analysis, made from the exponential of the original fractal
in order to give a log-normal density distribution function.  } 
\end{figure}
\clearpage

\begin{figure}
\plotone{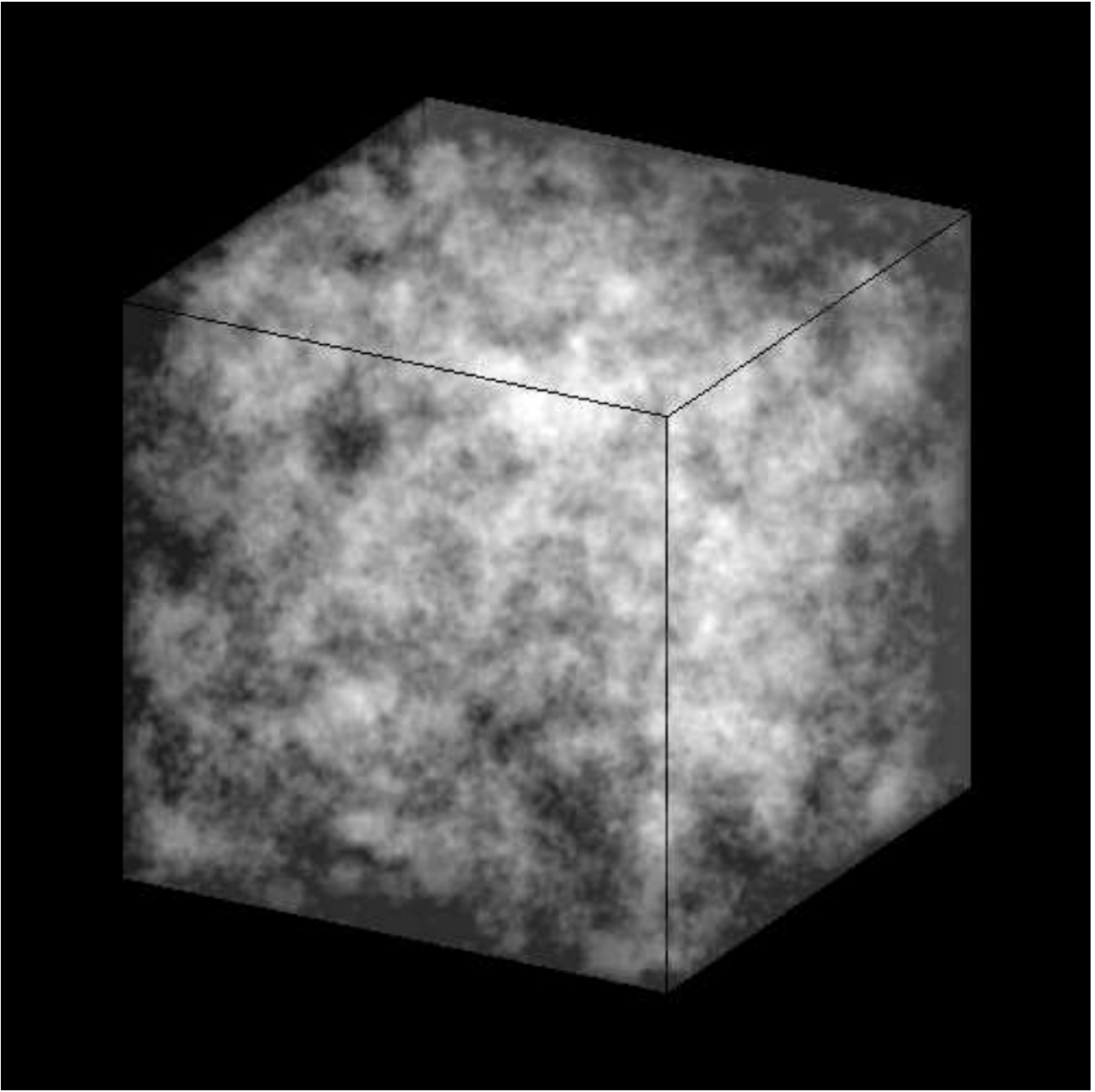}
\caption{Three-dimensional fractal model clipped at 0.03 times
the peak intensity. This type of model has the closest agreement
to interstellar gas, giving the observed volume filling
factor and mass spectrum for clumps.  } 
\end{figure}
\clearpage

\begin{figure}
\plotone{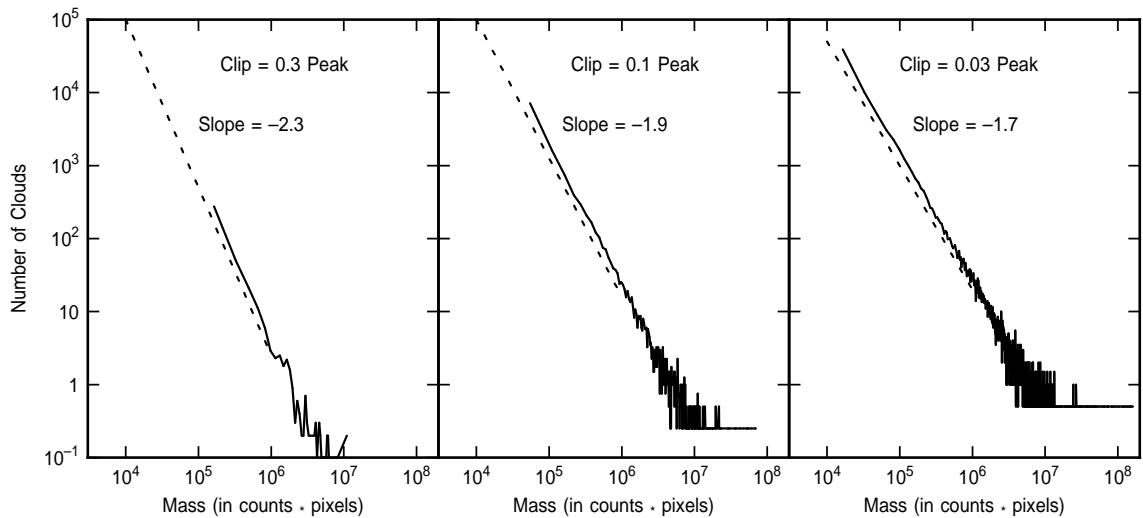}
\caption{Cloud mass spectra for three clipping levels relative to the
peak intensity.  These spectra are averages over 10, 4, and 2 random
model fractals, respectively, and so the minimum counts of 0.1, 0.25,
and 0.5 correspond to mass intervals in which only one model had a cloud.
The solid curves are the mass spectra and the dashed lines are indicative
of the slopes. Lower clipping levels have lower minimum masses and higher
maximum masses, but about the same number of clouds at a given mass.
The mass is from the sum of the density values in each cloud and is in
arbitrary units.  } 
\end{figure} 
\clearpage

\begin{figure}
\plotone{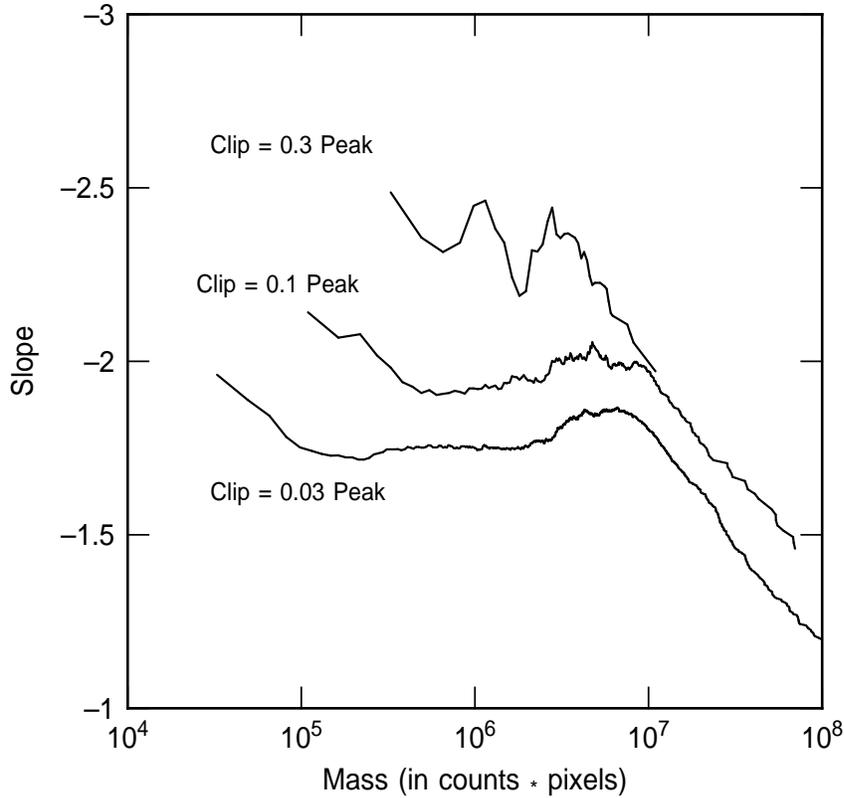}
\caption{Running average slopes of the
cloud mass spectra, averaged from the lowest mass up
to the mass plotted on the abscissa (in arbitrary units).
Results for the three clipping levels are shown.  The lowest
clipping level gives a power-law mass spectrum with the shallowest
slope, $\sim-1.7$. This shallow slope is 
similar to the observed slope for interstellar clouds and clumps,
and it is also from the clipping level that gives the same
filling factor as real clouds. The steeper slopes for higher
clipping levels might be more appropriate for star clusters and
mm-continuum sources, which sample denser regions of clouds.
}
\end{figure}

\clearpage

\begin{figure}
\plotone{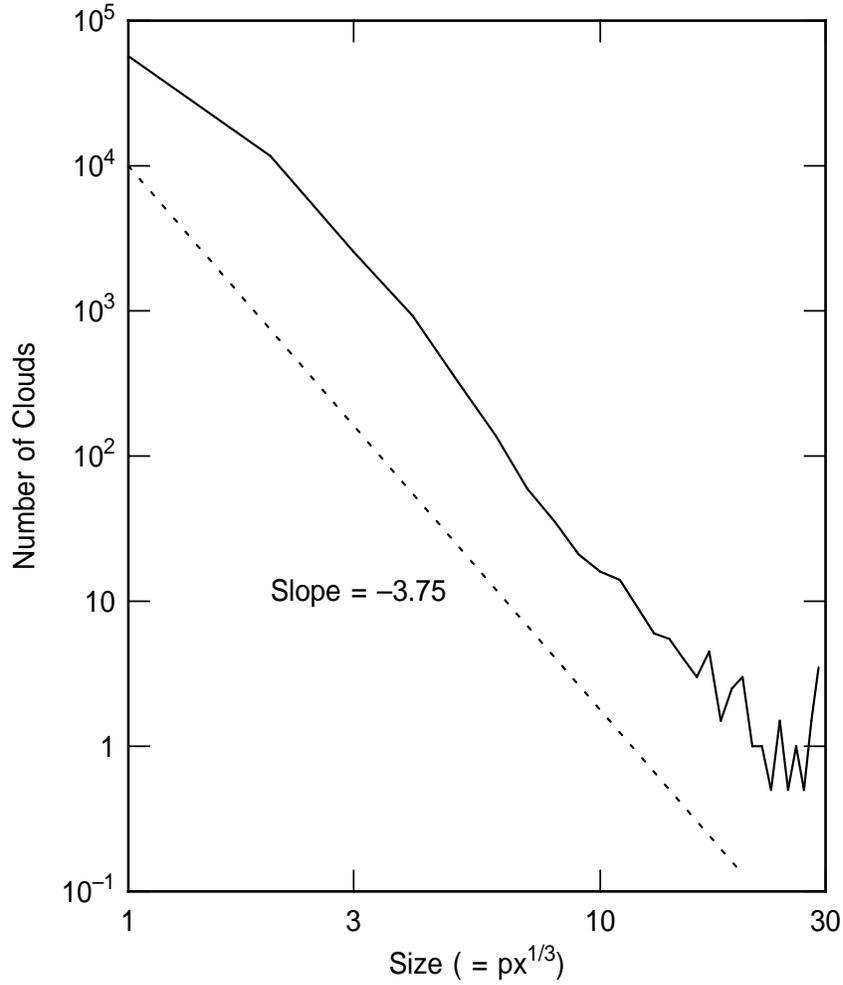}
\caption{Size spectrum for clouds in the model clipped at 0.03 
times the peak.  The power law nature of the cloud
size spectrum leads to the power law mass spectrum
even though the density distribution function is not a power 
law, but a log-normal.  The slope of the size spectrum can be used to
estimate the average fractal dimension.
}
\end{figure}
\clearpage

\begin{figure}
\plotone{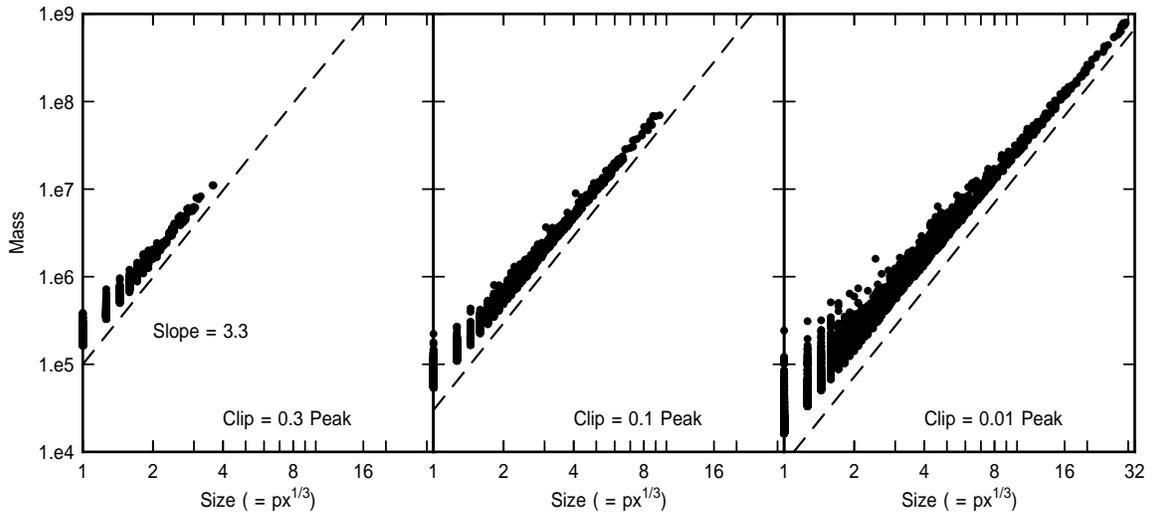}
\caption{Mass versus size for clouds found in the fractal models.
The clouds have a nearly constant density that scales with the
clipping value used to define them. 
}
\end{figure}

\begin{figure}
\plotone{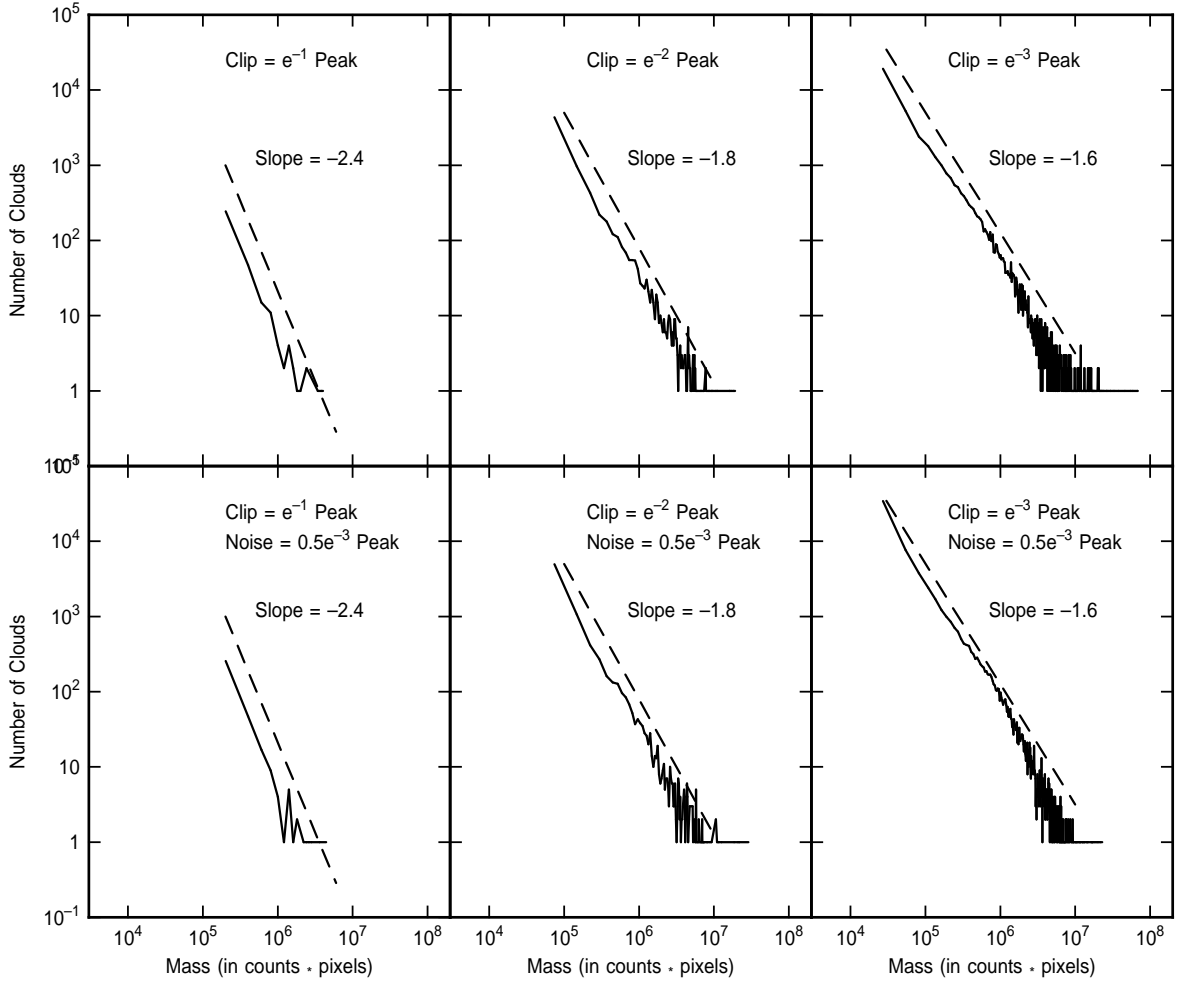}
\caption{Cloud mass spectra for three clipping levels relative
to the peak intensity, as in Fig. 3, but now with clouds
defined by the resolved peaks, rather than by 
all of the connected emission
above the clipping level. 
The top panels are without noise, the bottom panels are for
a fractal with noise
added at the level of $\pm0.5e^{-3}$ times the peak density. 
Only single models are shown here, not averages of
several models as in Fig. 3. 
The mass is in arbitrary units.}
\end{figure}

\begin{figure}
\plotone{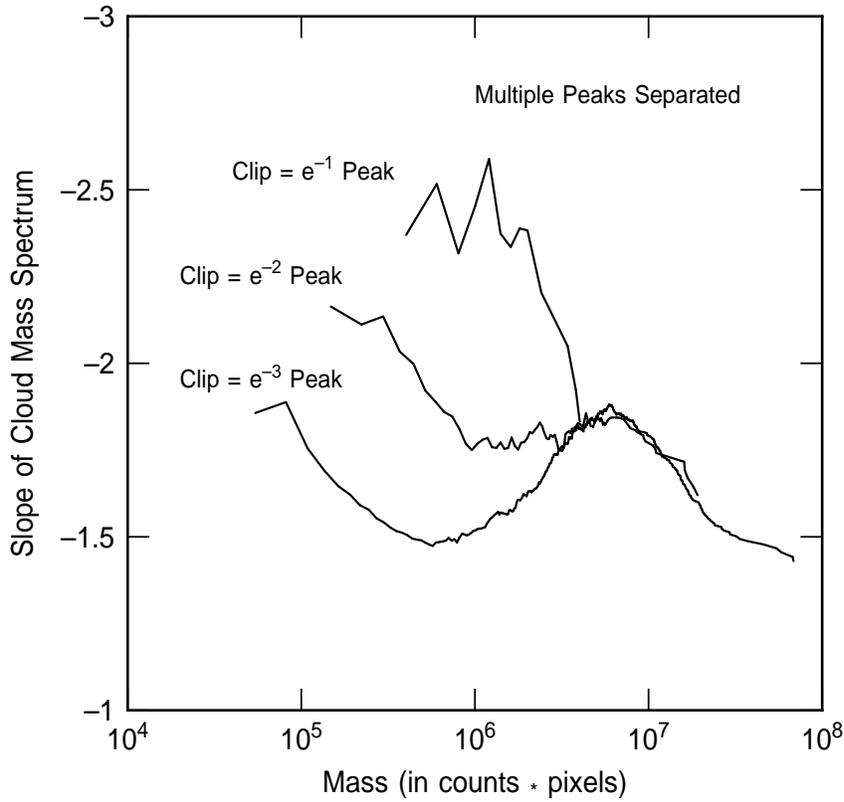}
\caption{Running average slopes of the
noise-free
cloud mass spectra shown at the top of
Fig. 7, averaged from the lowest mass up
to the mass plotted on the abscissa (in arbitrary units).
Results for the three clipping levels are shown, as in 
Fig. 4, but now with clouds defined by the resolved peaks. 
}
\end{figure}

\end{document}